\documentclass{iopconfser} 
\newcommand{\text}{\rm}

\usepackage{graphicx}
\usepackage{epsfig}
\usepackage{natbib}
\usepackage{har2nat}

\usepackage{amsmath}
\usepackage{amssymb}
\usepackage{lineno}

\usepackage{wrapfig}

\begin{document}

\title{On the Influence of Parallax Effects in Thick Silicon Sensors in Coherent Diffraction Imaging}

\author{Markus Kuster$^{1}$, Robert Hartmann $^{2}$, Steffen Hauf$^{1}$, Peter Holl$^{2}$, Tonn R\"uter$^{1, \ddagger}$, Lothar Str\"uder$^{2}$\vspace{-1cm}}
\email{markus.kuster@xfel.eu}

\affil{$^{1}$ European XFEL GmbH, Holzkoppel 4, 22869 Schenefeld, Germany}
\affil{$^{2}$ PNSensor GmbH, Otto-Hahn-Ring 6, 81739 M\"unchen, Germany}
\affil{$^{\ddagger}$ on leave}

\begin{abstract}
    Structure determination of, e.g., single molecules is a key application of X-ray free-electron lasers (XFELs)  and $4^\text{th}$ generation synchrotron sources, particularly using the coherent and pulsed X-ray radiation from XFELs. Scientific interest focuses on understanding the physical, biological, and chemical properties of samples at the nanometer scale. The X-rays from XFELs enable Coherent X-ray Diffraction Imaging (CXDI), where coherent X-rays irradiate a sample, and a far-field diffraction pattern is captured by an imaging detector.
    
By the nature of the underlying physics, the resolution, at which the molecular structure of a sample can be probed with the CXDI technique, is limited by the wavelength of the X-ray radiation and the exposure time if a detector can record the diffraction pattern up to very large scattering angles. The resolution that can be achieved under real experimental conditions, depends strongly on additional parameters. The Shannon pixel size, linked to the detector resolution, the coherent dose that can be deposited in the sample without changing its structure, the image contrast, and the signal-to-noise ratio of the detected scattered radiation at high $q$, i.e., at high scattering angles 2$\Theta$, have a strong influence on the resolution. The signal-to-noise ratio at high $q$ defines the “effective” maximum solid angle in a specific experiment setup at which a detector can efficiently detect a signal and in consequence determines the achievable resolution. The image contrast defines how well bright image features can be distinguished from dark ones. We present the preliminary results of our study on the influence of the point spread function on the signal-to-noise ratio, image contrast, position resolution, and achievable sample resolution for different pixel sizes.
\end{abstract}

\section{Introduction}
\label{Sec:introduction}

The continuous improvement of XFELs and $4^\text{th}$ generation synchrotron sources in the last decade has driven the need for high-performance imaging detectors to fully leverage their capabilities. Detectors at XFEL and synchrotron facilities are often designed for specific scientific applications and optimized accordingly. Looking into the future, new detector technologies must evolve to match the advancing performance of these light sources. The next generation will need enhanced sensitivity, frame rate, and spatial resolution to support ongoing upgrades and increasingly complex experiments. As scattering experiments push spatial resolution boundaries, next-generation detectors will be essential for probing the $q$ space with spatial resolution better than $100\,\mu\text{m}$.

Coherent X-ray diffraction imaging (CXDI) has become a powerful tool for imaging, e.g., nanoscale biological structures, utilizing coherent beams from Free Electron Lasers. The achievable resolution in CXDI is primarily limited by the underlying physics, i.e., the X-ray wavelength and exposure time, provided that a detector can capture diffraction patterns at very large scattering angles. Real experimental conditions introduce additional factors, such as the Shannon pixel size and the coherent dose that can be deposited in the sample without altering the sample structure. The signal-to-noise ratio (SNR) at high $q$ determines the maximum solid angle a detector can efficiently detect a signal and in consequence the achievable resolution \citep[see, e.g.,][]{Starodub:2008a}.

Planar, large-area photon imaging pixel detectors are commonly used in diffraction experiments at XFELs. Depending on detector parameters and geometry, the spherical nature of scattering physics can cause significant geometric distortion of the diffraction image at scattering angles beyond $10^\circ$. \citet[][]{Huelsen:2005a} observed signal shifts for the PILATUS~1M detector ($217\,\mu\text{m}$ pixel size) ranging from $0.05$ to $0.70$ times the pixel size at scattering angles between $10^\circ$ and $70^\circ$. With photon counting detectors like the PILATUS~1M these aberrations can be corrected using clustering algorithms \citep[see, e.g.,][]{Ihle:2017a}, while for charge-integrating detectors similar corrections face greater challenges due to systematic uncertainties and limitations inherent in the method to be used (e.g., deconvolution techniques). Distortions, such as the parallax effect in thick silicon sensors, may reduce the detector's position resolution, the SNR, and the image contrast at high $q$ values and limit the resolution of a scattering experiment.
\begin{figure}
    \begin{center}
        \includegraphics[width=0.80\columnwidth]{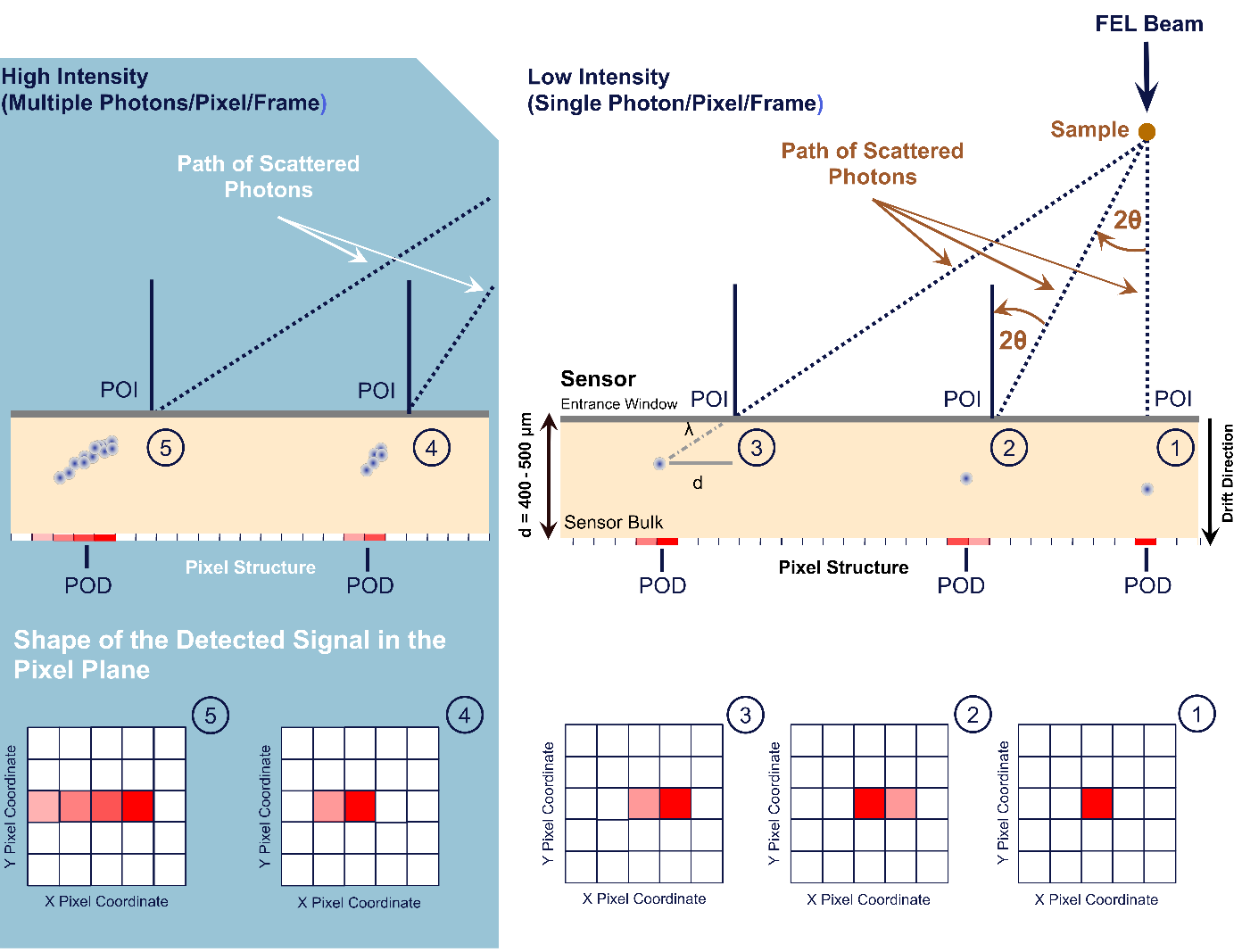} 
    \end{center}
    \label{Fig:schematic-view-parallax-effect}
    \caption{Illustration of the geometry of a CXDI scattering experiment (top panel) using a planar detector, the location of the POI and the POD (top panel) for different scattering angles and intensities, and the resulting signal distribution in the pixel plane (bottom panel). A darker red color means that more charge was collected in the pixel. Please note that the dimensions are not to scale.}
\end{figure}
In this work, we present preliminary results on how parallax and other geometric effects in a planar integrating detector with a $500\,\mu\text{m}$ thick silicon sensor and different pixel sizes impact the SNR, contrast, position resolution, and overall achievable CXDI experiment resolution.

\section{Scattering and Detection Geometry}
\label{Sec:experiment-geometry}
The typical geometry for a CXDI scattering experiment is illustrated in Fig.~\ref{Fig:schematic-view-parallax-effect}. The sample to be studied is irradiated by a focused XFEL beam, and the scattered radiation is detected by a position-sensitive imaging detector. Photons scattered at an angle $2\Theta$ hit the sensor surface at the point of incidence (POI) with an angle of incidence (AOI) equal to $2\Theta$. Elastically scattered photons from the sample, which are mono-energetic, are absorbed in the sensor bulk after the distance $\lambda$, generating a charge cloud that drifts in the electric field $E$ toward the readout anode (pixel structure). The charge cloud widens during the drift, reaching a diameter of tens of $\mu\text{m}$ driven by Coulomb repulsion and ambipolar diffusion.

In the case of single-photon detection, where less than one photon per pixel is absorbed during one integration time, the charge distribution's shape after the drift process at the pixel structure is shown in the bottom panel of Fig.~\ref{Fig:schematic-view-parallax-effect} for the exemplary cases $1$ to $3$. When more than one photon is absorbed per integration cycle, the charge clouds generated by the individual photons overlap and can create a more extended signal structure at the pixel plane, as illustrated in sketches $4$ and $5$.
\begin{figure}[t]
    \begin{center}
        \includegraphics[width=0.9\columnwidth]{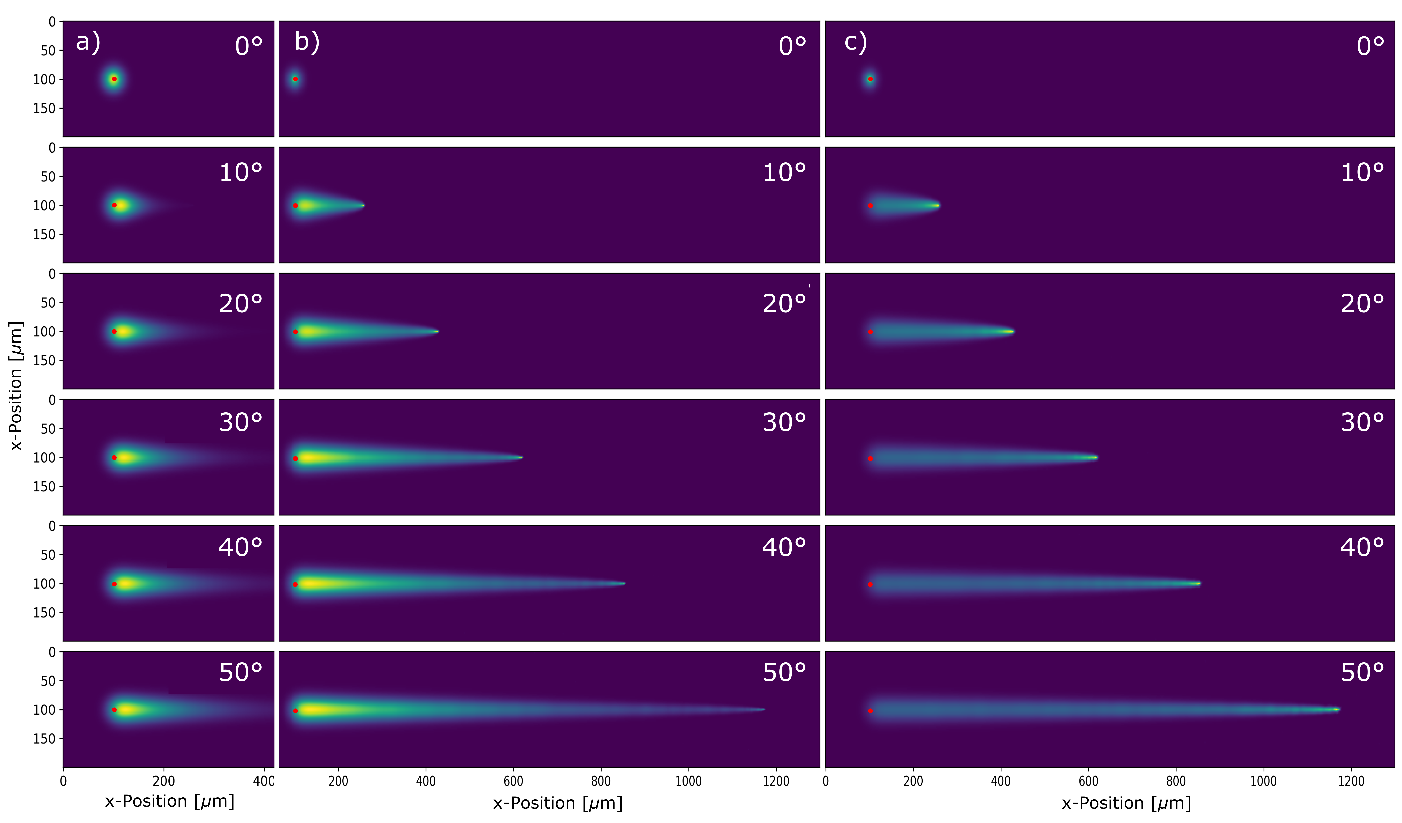}     
    \end{center}
    \caption{The simulated shape of the PSF depending on the incident photon energy and scattering angle $2\Theta$ is shown. The resolution of the PSF distributions is $1\,\mu m$. The left panel a) shows the PSF for $8\,\text{keV}$ photons, the middle panel b) for $12\,\text{keV}$ and the right panel c) for $20\,\text{keV}$ photons with increasing scattering angle from top to bottom starting from $0^\circ$  to $50^\circ$. The red dot located at the pixel coordinates $x=100\,\mu m$ and $y=100\,\mu m$ marks the photon's POI for each case.}
    \label{Fig:PSF-vs-Energy-AOI}
\end{figure}

A single photon impacting perpendicularly to the entrance window (case 1) creates a signal in the pixel directly below the POI, provided the charge cloud's diameter is smaller than the pixel pitch and completely remains contained within the pixel volume. This results in the point of detection (POD) coinciding with the POI within one pixel's precision. However, as the scattering angle $2\Theta$  increases, the POD shifts further from the POI, with the displacement $d=|x_\text{POD} - x_\text{POI}|$ proportional to $\sin(2\Theta)$, as shown in cases 2 and 3. These cases additionally depict situations where the charge cloud intersects a pixel boundary, depositing charge in two neighboring pixels. In case 2, most charge is collected in the left pixel (dark red), while in case 3, it's collected in the right pixel. At high scattering intensities, multiple photons can deposit their energy in the sensor bulk during one image integration cycle, leading to overlapping charge clouds that are collected by one or several pixels, making it impossible to differentiate contributions from individual photons, reconstruct their POI, or correct individually for the displacement $d$. The measured signal's topology is influenced by the photon energy and scattering angle. The following discussion will focus on this "High Intensity" scenario.

\section{Modeling the Spatial Response - Point Spread Function}
\label{Sec:Modeling the spatial response}
For the mathematical description of the spatial response of a planar pixelated detector in a scattering experiment, we use the generalized concept of the Point Spread Function (PSF). We consider the response of the sensor to photons with energy $E$ incident at an AOI of $2\Theta$. The intensity distribution of a pointlike object as detected by the detector $I(x,y)$ is then given by
\begin{equation}
    \label{Eq:PSF-Definition}
    I(x,y) = \int_{-\infty}^{+\infty}\int_{-\infty}^{+\infty} I_0(u,v)\, PSF(x-u,y-v,E,2\Theta)\,dudv
\end{equation}
where $I_0$ is the scattered intensity in units of photons per X-ray pulse and $u,v$ the integration variables of this convolution integral. The point spread function $PSF(x,y,E,2\Theta)$ depends on the location of the POI in detector pixel coordinates $(x,y)$, the photon energy $E$, and the scattering angle $2\Theta$. To model the relevant PSF parameter space and study its influence on CXDI applications, we simulated X-ray absorption in the sensor bulk for photon energies ranging from $E=6\,\text{keV}$ and $25\,\text{keV}$ in $\Delta E= 1\,\text{keV}$ steps. Using a quadratic pixel geometry with a $1\,\mu\text{m}$ pixel pitch provides sufficiently high resolution, allowing us to infer the effects of larger pixel sizes by re-binning the PSF. 
\begin{wrapfigure}{l}{0.49\columnwidth}
    \vspace{-0.6cm}
    \includegraphics[width=0.48\columnwidth]{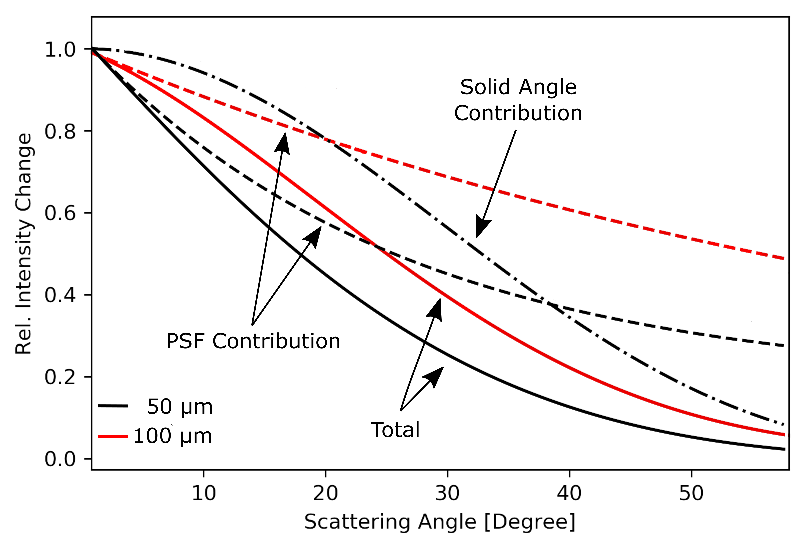}
    \vspace{-0.2cm}
    \caption{The total relative change in signal measured in one pixel as a function of scattering angle for $12\,\text{keV}$ photons, comparing two pixel sizes: $50\,\mu\text{m}$ (black solid line) and $100\,\mu\text{m}$ (red solid line). The intensity drop due to the decreasing solid angle with increasing scattering angle is calculated according to Eq.~\ref{Eq-Solid-Angle-Precise} (dash-dotted black line). Additionally, the simulated influence of the parallax effect as a function of scattering angle $2\Theta$ is shown by the dashed lines (see Section~\ref{Sec:signal-to-noise}).}
    \vspace{-0.5cm}
    \label{Fig:parallax-PSF-solid-angle-correction}
\end{wrapfigure}
Our simulation treats X-ray absorption, the subsequent charge transport, and collection processes in the fully depleted sensor bulk separately, enabling us to distinguish their contributions to the PSF. We model the X-ray electromagnetic interaction using the Geant4 toolkit Version 10 \citep{geant4,Allison:2006a}. We are using the Livermore physics list to simulate photon and electron interaction. For each AOI, the sensor was illuminated with $10^5$ photons striking the entrance window at the POI, located at $x_{\text{POI}}= 100\,\mu\text{m}$ and $y_{\text{POI}} =100\,\mu\text{m}$  (marked with a red dot) as shown in the coordinate systems in Fig.~\ref{Fig:PSF-vs-Energy-AOI}.

The resulting normalized PSFs for four different photon energies and six AOIs are shown in  Fig.~\ref{Fig:PSF-vs-Energy-AOI} on a logarithmic color scale. As evident, for the perpendicular incidence ($2\Theta = 0^\circ$), the PSF diameter decreases with increasing photon energy. This aligns with expectations, as higher-energy photons are absorbed deeper in the sensor bulk, reducing the drift distance and time for the charge cloud to spread due to diffusion and Coulomb repulsion. As the angle of incidence exceeds $10^\circ$, the PSF adopts an elongated drop-like shape with a narrow tail from photons likely absorbed deeper in the sensor's bulk. Fluorescent X-rays or kinetic electrons scattered in the sensor material contribute to the weak tail far from the primary POI. As the photon energy increases, their penetration depth increases, and the photons are most likely absorbed close to the pixel structure. As a consequence the PSF maximum moves away from the POI to the tail of the PSF (compare Fig. \ref{Fig:PSF-vs-Energy-AOI} panels b) and c) for, e.g., $2\Theta=40^{\circ}$). 

\begin{figure}
    \begin{center}
        \includegraphics[width=0.85\columnwidth]{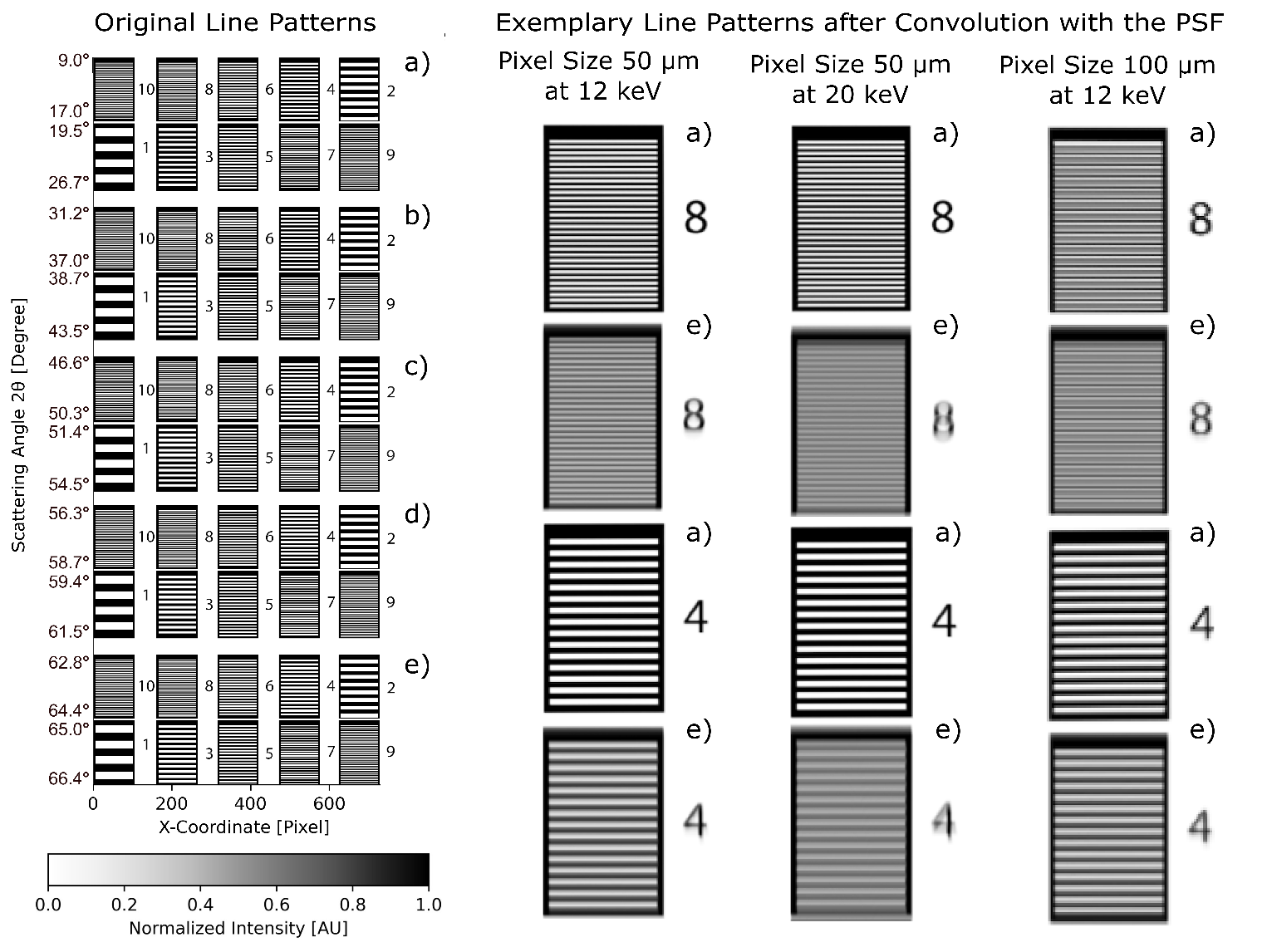} 
    \end{center}
    \vspace{-0.8cm}
    \caption{On the left side the original line density patterns used as input for our simulations depending on the scattering angle $2\Theta$ are shown. On the right exemplary convoluted patterns 4 and 8 are shown for $50\,\mu\text{m}$ pixel size and $12\,\text{keV}$ photon energy (left), $50\,\mu\text{m}$ pixel size and $20\,\text{keV}$ photon energy (middle), and $100\,\mu\text{m}$ pixel size and $12\,\text{keV}$ photon energy (right), and the scattering angle range a) and e). The lookup table is valid for all patterns shown. For a detailed description of these results, we refer the reader to the text.}
    \vspace{-0.3cm}
    \label{Fig:simulated-line-density-panels}
\end{figure}

\section{Signal to Noise, Contrast and Spatial Resolution}
\label{Sec:signal-to-noise}
The elongated shape of the PSF, leading to a redistribution of the incident scattered intensity $I_0$ to a larger number of pixels at increasing scattering angles $2\Theta$ and photon energies $E$, can affect spatial resolution, per-pixel SNR, image contrast, and the POD. Consequently, the detected intensity per pixel area is a function of the scattering angle and photon energy. To quantify this effect, we convolved a delta-peak like $I_0$ with the PSF as a function of $2\Theta$, where each peak represents 10,000 photons. We assumed a planar, gapless detector with $50\,\mu\text{m}\times50\,\mu\text{m}$ or $100\,\mu\text{m}\times100\,\mu\text{m}$ large pixels, illuminated by 12 keV photons. The maximum value of the resulting intensity distribution $I(x,y)$ as a function of $2\Theta$ is shown in Fig.~\ref{Fig:parallax-PSF-solid-angle-correction} as a black dashed line for $50\,\mu\text{m}$ pixels and a red dashed line for $100\,\mu\text{m}$ pixels, labeled 'PSF Contribution'.

Another contribution is the decreasing solid angle with increasing scattering angle. The solid angle extended by one pixel with the area $A$ depends on the distance between the sample and pixel's center $R$ and is \cite{Grillo:2008a}
\begin{equation}
    \label{Eq-Solid-Angle-Precise}
    \Omega (2\Theta) = \frac{A}{R}\, \cos^2(2\Theta)\, \cos(2\Theta).
\end{equation}
The decreasing solid angle reduces the intensity per pixel by $\approx88\%$ at $2\Theta = 60^\circ$. When all contributions are combined, the relative signal drops from $100\%$ at the sensor's center to below $10\%$ at its edge, at $2\Theta = 60^\circ$ and so does the SNR. Figure~\ref{Fig:parallax-PSF-solid-angle-correction} shows the progression of relative intensity per pixel area as a function of the scattering angle, along with the contributions of the PSF and the solid angle. For this exemplary case, the PSF is the dominant factor of the overall effect for $2\Theta\gtrsim 40^\circ$.

The resolution of a sensor depends on more than just its ability to separate the smallest feature of interest. Factors such as PSF blurring, geometric distortions, and contrast variations also play a role. To quantitatively estimate the impact of the PSF on spatial resolution and image contrast, we used line patterns with defined line densities, shown in Fig.~\ref{Fig:simulated-line-density-panels}. By design, the line density doubles as the pattern number doubles, with patterns grouped into panels labeled a) to e) and numbered 1 to 10. To simulate different scattering angles from $\approx 9^\circ$ to $\approx 66^\circ$, the panel was convolved with the PSF for each angle. The result is shown in Fig.~\ref{Fig:simulated-line-density-panels}. The direct XFEL beam hits the detector plane perpendicularly at $y = 1024$ or $2048$, defining $2\Theta = 0$. $2\Theta$ increases with decreasing $y$-values.

\begin{figure}
    \includegraphics[width=0.45\columnwidth]{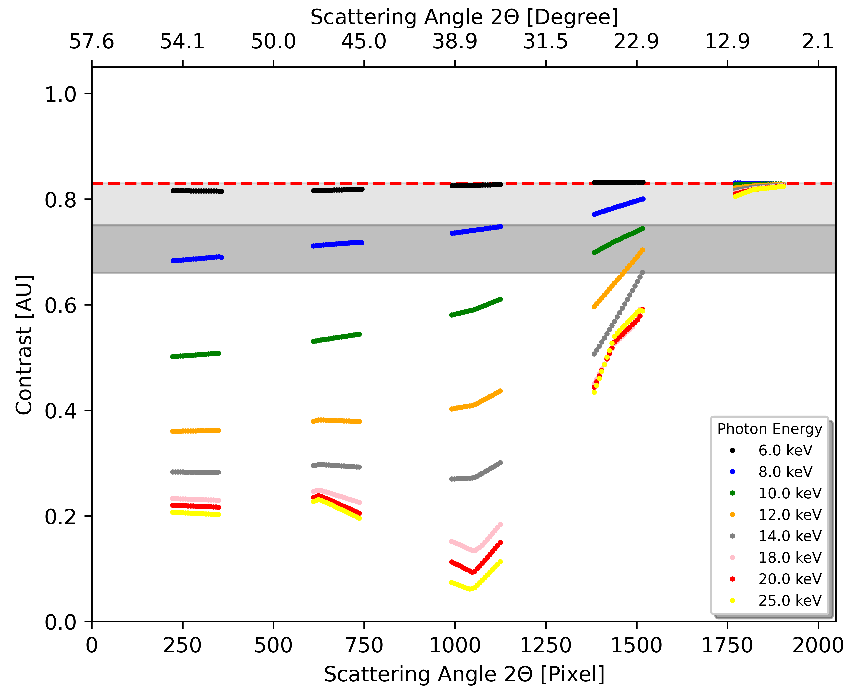}
    \hfill
    \includegraphics[width=0.45\columnwidth]{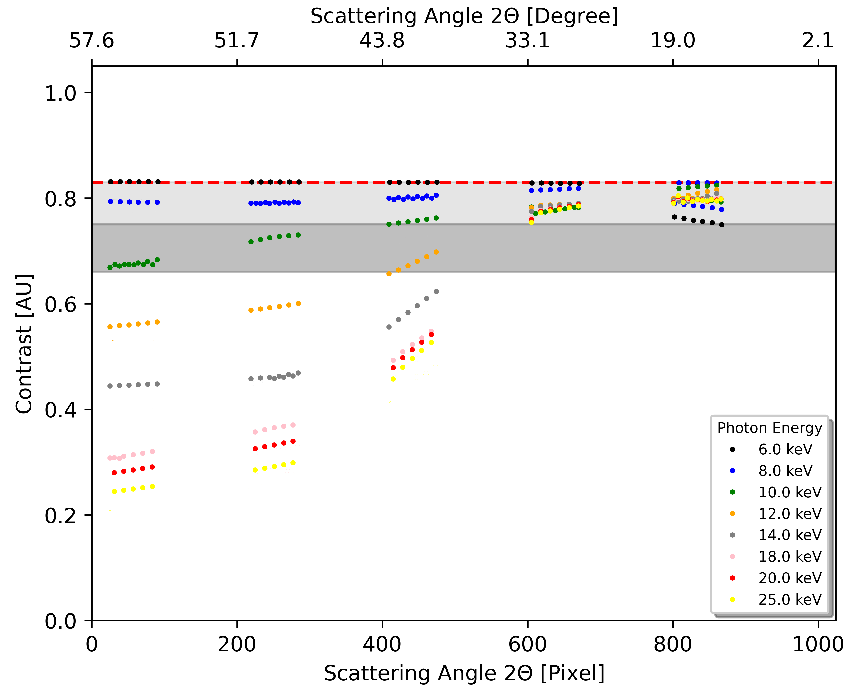}
    \caption{Left: Simulated radial dependence of the contrast as a function of scattering angle $2\Theta$ and photon energy for pixel sizes of $50\,\mu\text{m}\times50\,\mu\text{m}$ (left) and $100\,\mu\text{m} \times 100\,\mu\text{m}$ (right). The detector center is located at $2\Theta=0^\circ$, corresponding to pixel coordinates $x=2048$ for the smaller pixels, and $x=1024$ for the larger pixels.}
    \label{Fig:contrast-vs-scattering-angle-energy}
    \vspace{-0.5cm}
\end{figure}

As expected, the line patterns become increasingly blurred as the scattering angle increases. The blurring intensifies with higher photon energies and smaller pixel sizes. For instance, pattern $4$ at $20\,\text{keV}$ and a $50\,\mu \text{m}$ pixel size shows strong contrast at $2\Theta$ between $9^\circ$ and $17^\circ$ (panel a), but the contrast significantly diminishes at $2\Theta > 62.8^\circ$ (panel e). These line patterns were used to quantify the contrast 
\begin{equation}
    \label{Eq-Contrast}
    K = \frac{L_{\text{max}}-L_{\text{min}}}{L_{\text{min}}}
\end{equation}
based on scattering angle and photon energy, where $L_{\text{max}/\text{min}}$ is the maximum and minimum brightness of the line pattern, calculated for a specific range of scattering angles. The result is shown in Fig.~\ref{Fig:contrast-vs-scattering-angle-energy} for two pixel sizes $50\,\mu\text{m}$ (left image) and $100\,\mu\text{m}$ (right image). For smaller pixels, the contrast drops below the initial value of $0.86$ (at $2\Theta = 0$, red dashed line) by more than 10\% (light grey area) or 20\% (dark grey area) starting at $2\Theta\gtrapprox 20^\circ$. As photon energy increases, the contrast decrease is faster and more pronounced, falling below $40\%$ for $E \gtrapprox 12\,\text{keV}$.

\section{Influence on the CXDI Resolution}
\label{Sec:CXDI-resolution}

\citet{Starodub:2008a} and further work referenced by these authors \citep[e.g.,][]{Howells:2009a, Shen:2004a} demonstrated that for reconstructing a sample's structure using CXDI with a target resolution $d$, it is necessary to detect a statistically significant signal at the detector pixel corresponding to $q_{\text{max}}=2\pi/d$. For a detector with $N \times N$ pixels extending to $q_{\text{max}} = N \frac{\Delta q}{2}$, the authors derived a theoretical relation for the counts $P$ detected during a time interval $\Delta t$ in a pixel corresponding to  the resolution $d$:
\begin{equation}
    \label{Eq:relation-countrate-CXDIresolution}
    P = \frac{1}{8\pi^2s^2}\,r_e^2\lambda^2d^4 |\rho|^2 I_0 \Delta t\,.
\end{equation}
Here, $\rho =\sum_i \,n_{ai}\,(f_{1i}+if_{2i})$ is the effective complex electron density, $n_{ai}$ the atomic concentrations in the sampley, $\lambda$ the wavelength, $I_0$ the incident X-ray flux, $f_{1i}$ and $f_{2i}$ the atomic scattering amplitudes for the $i$th type of atom, and $\Delta t$ the integration time. Solving Eq.~\ref{Eq:relation-countrate-CXDIresolution} provides an estimate of the achievable resolution based on the detected intensity at $q_{\text{max}}$. Assuming successful 3D reconstruction of the sample is just possible when observing a very weak signal just above the detection limit at $q_{\text{max}}$, incorporating the influence of PSF suggests a reduction in achievable resolution by up to a factor of $\approx 2.7$ for a planar detector covering $2\Theta = 60^\circ$ at $12\,\text{keV}$ with $50\,\mu\text{m}$ pixels.

\section{Conclusions and Outlook}
\label{Sec:conclusions-and-outlook}
The quality of diffraction images is affected by the parallax effect inherent in large-area planar integrating detectors with thick silicon sensors. Minimization of PSF effects can be achieved, for example, by optimizing the detection geometry from planar to curved when optimal $q$ space sampling, achievable by smaller pixels, is required. By choosing the sensor design, the charge drift time and thus the lateral width of the PSF can be reduced. To some extent, the observed effects can be partially corrected using deconvolution techniques during data processing. With a planar detector, photons are detected at a POD that is significantly displaced from the POI, with the displacement being on average proportional to the sensor thickness and $\sin(2\Theta)$ at high scattering angles. The non-pointlike PSF redistributes scattered intensity across multiple pixels, causing a decrease in measured intensity per pixel, SNR, image contrast, and image blurring as the scattering angle $2\Theta$ increases, pixel size decreases, and photon energy rises. Our preliminary results show that for $100\,\mu\text{m}$ pixels, contrast decreases by less than $10\%$ up to $25\,\text{keV}$ and $2\Theta \lessapprox 30^\circ$. For smaller pixels, contrast drops to $50\%$ at $2\Theta = 24^\circ$ and $14\,\text{keV}$. For a planar detector with $50\,\mu\text{m}$ pixels covering $2\Theta = 60^\circ$ at $12\,\text{keV}$, sample resolution is reduced by up to a factor of $2.7$ compared to an image in which a significant signal can just be detected at the highest scattering angle the detector covers. To preserve image contrast and resolution from $6$ to $25\,\text{keV}$, the angular coverage of a planar detector should not exceed $\approx13^\circ$ for $50\,\mu\text{m}$ pixels and $\approx35^\circ$ for $100\,\mu\text{m}$ pixels. 

\appendix



\bibliography{mnemonic,detback,parallax}

\end{document}